\documentclass[a4paper,11pt]{article}
\usepackage{pos}
\usepackage{lineno}
\usepackage{float}
\newcommand{\etaphi}{\ensuremath{(\eta,\phi)}~}

\newcommand{\ptJet}{\ensuremath{p_{\mathrm{T, jet}}}}

\newcommand{\ptJetRange}[2]{\ensuremath{#1 \leq p_{\mathrm{T, jet}} < #2~\mathrm{GeV}/c}}

\newcommand{\ptJetGeq}[1]{\ensuremath{p_{\mathrm{T, jet}} \geq #1~\mathrm{GeV}/c}}

\newcommand{\ptTrack}{\ensuremath{p_{\mathrm{T, track}}}}

\newcommand{\ptAssoc}{\ensuremath{p_{\mathrm{T, assoc}}}}

\newcommand{\pp}{\ensuremath{p + p}~}

\newcommand{\g}{$g$}

\newcommand{\lesub}{\ensuremath{\rm LeSub}}

\newcommand{\Gevc}{\ensuremath{~\mathrm{GeV}/c}}

\newcommand{\EtTower}{\ensuremath{E_{\mathrm{T, tower}}}}

\newcommand{\rhoR}{\ensuremath{\rho({\mathbf{r}})}}

\newcommand{\sGev}{\ensuremath{\sqrt{s} = 200\ \mathrm{GeV}}}

\newcommand{\sGevPbPb}{\ensuremath{\sqrt{s_{_{\rm NN}}} = 2.76\ \mathrm{TeV}}}

\newcommand{\ptd}{\ensuremath{\rm p_{\mathrm{T}}^D}}

\newcommand{\qqToqq}{\ensuremath{qq\rightarrow qq}~}

\newcommand{\ggTogg}{\ensuremath{gg\rightarrow gg}~}

\newcommand{\addTwoPanelFig}[3]{
    \begin{figure}[H]
        \centering
        \includegraphics[width = 0.7\linewidth]{#1}
        \caption{#2}
        \label{#3}
    \end{figure}
}

\newcommand{\addTwoPanelInFig}[1]{\includegraphics[width = 0.68\linewidth]{#1}}

\newcommand{\addOnePanelInFig}[1]{\includegraphics[width = 0.49\linewidth]{#1}}

\title{Generalized angularities and differential jet shapes measurements from STAR at $\sqrt{s} = $ 200 GeV}

\author*[a]{Tanmay Pani}
\onbehalf{(for the STAR collaboration)}

\affiliation[a]{Rutgers University,\\
  136 Frelinghuysen Road, Piscataway, USA}

\emailAdd{tp543@physics.rutgers.edu}

\abstract{Jets from early stages of heavy-ion collisions undergo modified showering in quark-gluon plasma (QGP) relative to vacuum due to jet-medium interactions, which can be measured using observables like differential jet shape and generalized angularities. Differential jet shape (\rhoR) encodes radially differential information about jet broadening and has shown an average migration of charged energy away from the axes of quenched jets from Pb+Pb collisions at the LHC. Measurements of generalized angularities in presence of the medium from Pb+Pb collisions at the LHC show harder, or more quark-like jet fragmentation relative to vacuum. Measuring these distributions in heavy-ion collisions at RHIC will help us further characterize jet-medium interactions in a phase-space region complementary to that of the LHC. 

In these proceedings, we present the first fully corrected measurements of \rhoR, jet girth (\g), momentum dispersion (\ptd) and momentum difference of leading and subleading constituent particles (\lesub) observables, using hard-core jets in \pp collisions at \sGev, collected by the STAR experiment. Finally, the data are compared with model calculations and the physics implications are discussed.
}

\FullConference{HardProbes2023\\
 26-31 March 2023 \\
 Aschaffenburg, Germany\\}

\begin{document}
\maketitle
\section{Introduction}\label{sect:Intro}

Hard scattered partons from early stages of high-energy hadron collisions undergo successive, small-angle fragmentations, and eventually appear in the final state as collimated sprays of hadrons called~\textit{jets}. In heavy-ion collisions, jets traverse the quark-gluon plasma (QGP) medium and are modified relative to a  \pp baseline. This is known as \textit{jet quenching}~\cite{Gyulassy_2004}. Therefore, jets are used as probes of QGP, containing information of interaction between hard partons and QGP medium. One way to access the quenching information is by studying intra-jet angular distribution of energy relative to the jet-axis through generalized jet angularities, calculated as:

\begin{equation}\label{eq:GenAngIntro}
	\lambda_\beta^\kappa=\sum_{\mathrm{const} \in \mathrm{jet}} \left(\frac{p_{\rm T, const}}{p_ {\rm T,  jet}}\right)^\kappa r(\mathrm{const, jet})^\beta ,
\end{equation}
where \ptJet~is the jet's total momentum, and $r(\mathrm{const, jet}) = \sqrt{(\eta_{\rm jet}-\eta_{\rm const})^2 + (\phi_{\rm jet}-\phi_{\rm const})^2} $ is the \etaphi distance of a constituent from the jet-axis. Parameters $\kappa$ and $\beta$ tune experimental sensitivity to hard and wide-angle radiation, respectively. $\lambda^1_\beta$s are infra-red and collinear (IRC) safe angularites~\cite{Larkoski_2014}, which probe the average angular spread of energy around the jet-axis. They are radial moments of the  jet's momentum profile, also known as differential jet-shape (\rhoR), given by,

\begin{equation}\label{eq:RhoRIntro}
	\rho(\mathbf{r}) = \lim_{\delta r \rightarrow0}\Big\langle\frac{1}{\delta r}\frac{\sum_{|\mathbf{r_{\rm const}} - \mathbf{r}|<\delta r/2}p_{\rm T, const}}{p_{\rm T,  jet}}\Big\rangle _{\rm jets} ,
\end{equation}
where  $\mathbf{r}_{\rm const} =  (\eta_{\rm const}-\eta_{\rm jet})\hat{\eta}+ (\phi_{\rm const}-\phi_{\rm jet})\hat{\phi}$, and it follows that, 
\begin{equation}
	\lambda_\beta^1 = \int_{\rm jet} r^\beta\rho(\mathbf{r}) d\mathbf{r} .
\end{equation}

The jet angularity based observables like jet-substructure measurements in Pb+Pb collisions at \sGevPbPb~at the LHC, have shown quenched jets, on average, have migration of charged energy away from their axis relative to a \pp baseline \cite{CMS_2014243} and possibly a survivor bias toward harder, quark-like fragmentation \cite{Alice2018}. Similar measurements using jets with lower \ptJet~at RHIC, will help understand jet-medium interactions in a complementary phase-space region to LHC.

In this proceeding, jet girth ($g = \lambda_1^1$), momentum dispersion $(\ptd  =\sqrt{ \lambda_0^2} )$ and the differential jet-shape ($\rho(\mathbf{r})$) are measured in \pp collisions \sGev~to set a baseline for heavy-ion collisions at  RHIC. We also calculate a non-angularity based jet observable \lesub~which gives a measure of the hardest splitting of the jet:
\begin{equation}\label{Eq:LeSubIntro}
	LeSub = p_{\rm T, constituent}^{\rm leading} - p_{\rm T, constituent}^{\rm subleading} .
\end{equation} 

\section{Dataset and Analysis Method}\label{sect:AnaMethod}
The analysis uses data from \pp collisions at \sGev~collected in 2012 using the Solenoidal Tracker At RHIC (STAR) detector system.~Charged-particle tracks and neutral energy depositions (towers) are measured using STAR's Time Projection Chamber (TPC) \cite{Anderson_2003} and Barrel Electromagnetic Calorimeter (BEMC) \cite{BEDDO2003725} detectors respectively. Together, they provide full azimuthal coverage with a pseudorapidity acceptance of $|\eta| \leq 1$.  The tracks and towers are clustered into jets using the anti-$k_{\rm T}$ algorithm with a jet resolution parameter $R = 0.4$, implemented using the FastJet library~\cite{Cacciari_2012}. To suppress contributions of fake tracks and combinatorial background (especially in the context of the larger  heavy-ion background), a ``hard-core'' constituent selection as was done in previous STAR analyses \cite{PhysRevC.105.044906} is applied, which only allows tracks (towers) with $c\ptTrack (\EtTower) \geq 2~\mathrm{GeV}$ to be clustered into jets. To enhance jet signal, only High-Tower (HT) triggered events, with at least one tower with $\EtTower \geq 4$ GeV are considered. After clustering, only jets completely falling within acceptance ($|\eta_{\rm jet}| \leq 0.6$) are kept. Jets with area, $A_{\rm jet} < 0.3$ are rejected to further reduce the fake jet contribution.

The distributions of \g, \ptd~and \lesub~are fully corrected for detector effects by using iterative bayesian unfolding, implemented using the RooUnfold library \cite{bennerRoounfold}. The unfolding requires a response matrix between particle-level and detector-level. This is constructed using an embedding simulation which involves PYTHIA-6 STAR tune \cite{adkins2019studying} events processed into detector hits using GEANT3 \cite{Brun:1987ma} and added to real zero-bias events from \pp collision environment. To calculate \rhoR, additional associated tracks not clustered into jets, but inside the jet cones are also used. This was done to look at the complete jet, around its hard core. Given a jet, tracks with $\ptAssoc \geq 1\Gevc$ and $r(\rm assoc, jet) \leq 0.4$ are used. The \rhoR~is corrected using bin-by-bin factors obtained from the aforementioned embedding simulation\footnote{Details of closure associated with the unfolding can be found in slides 25-33 in the talk associated with this proceeding, https: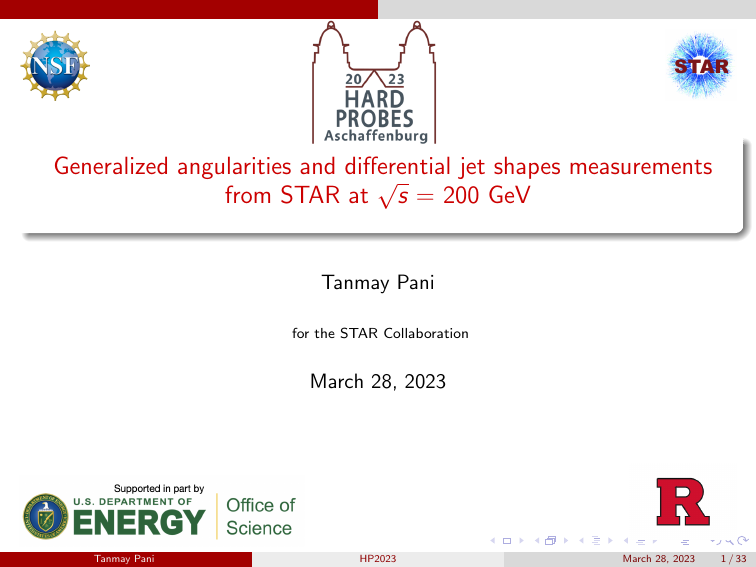}.

%

\section{Result and Discussion}\label{sect:Results}

\begin{figure}[H]
	\centering
	\addTwoPanelInFig{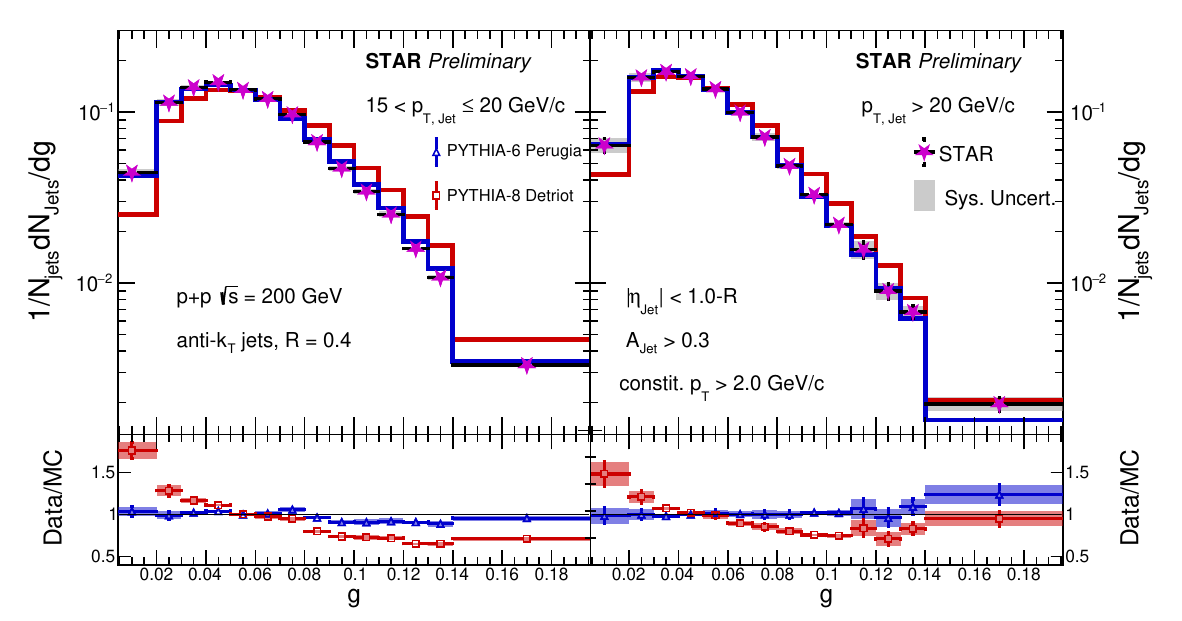}
	\addTwoPanelInFig{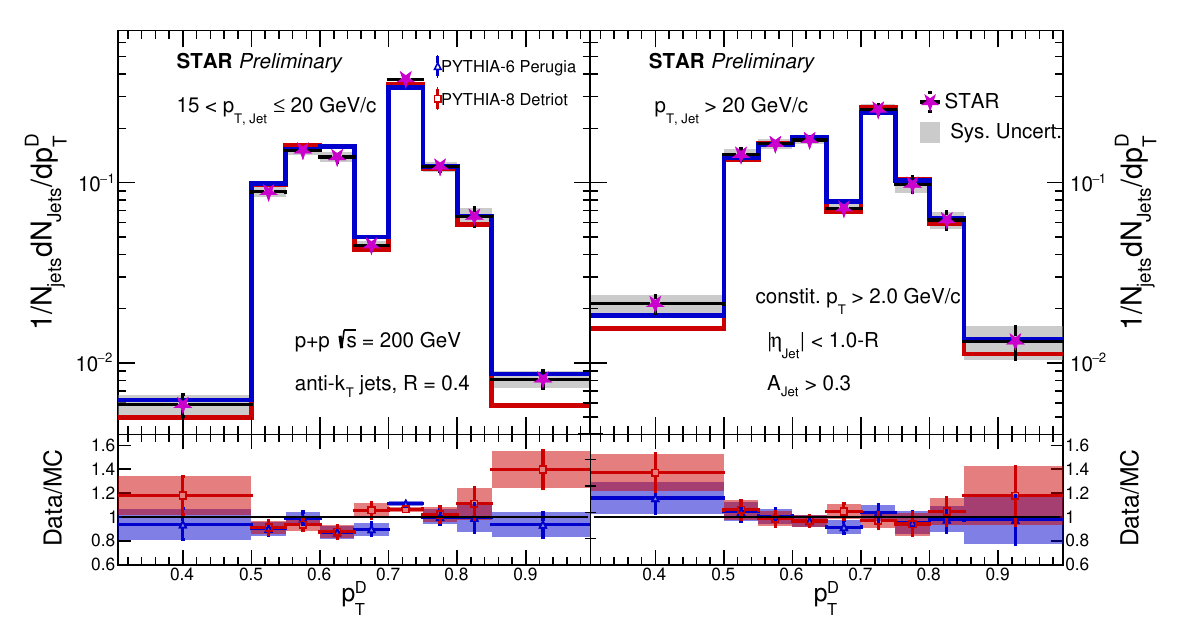}
	\addTwoPanelInFig{Lesub_Final.pdf}
	\caption{$g$ (top), \ptd (middle) and $LeSub$ (bottom) distributions (magenta stars, normalized to unity) for jets with \ptJetRange{15}{20} (left) and \ptJetGeq{20} (right). The results are compared to PYTHIA-6 (STAR) (blue) and PYTHIA-8 (Detroit) (red). The lower panels show the ratio of the data calculation to the PYTHIA-6 (STAR) (blue) and PYTHIA-8 (Detroit) (red).}
	\label{fig:JSOs}	
\end{figure}

Differential jet-shape as a function of $r = r(\rm assoc, jet)$ from the jet axis is shown in Fig.~\ref{fig:RhoR}. Girth ($g$), \ptd~and \lesub~distributions are shown in Fig.~\ref{fig:JSOs}. Systematic uncertainties are shown as shaded grey bands. On average, lower energy jets with \ptJetRange{15}{20} have higher $g$, lower $LeSub$ and more energy away from jet-axis than jets with \ptJetGeq{20}. 

\addTwoPanelFig{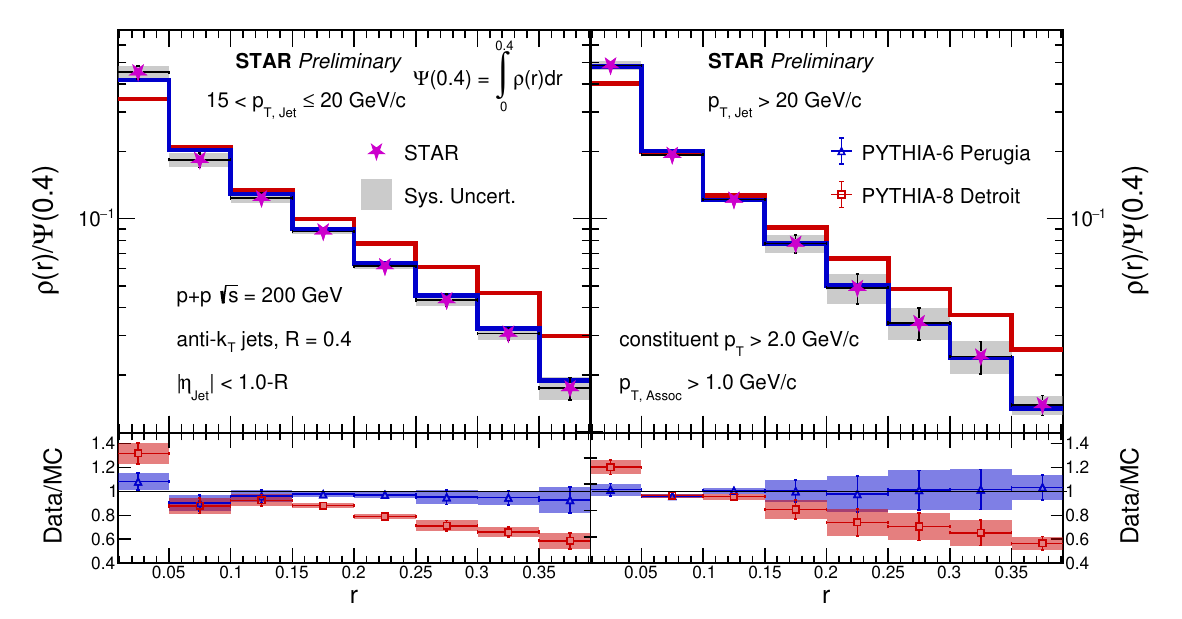}{\rhoR~vs $r$ (magenta stars, normalized to unity) for jets with \ptJetRange{15}{20} (left) and \ptJetGeq{20} (right). The results are compared to PYTHIA-6 (STAR) (blue) and PYTHIA-8 (Detroit) (red). The lower panels show the ratio of the data calculation to the PYTHIA-6 (STAR) (blue) and PYTHIA-8 (Detroit) (red).}{fig:RhoR}

\begin{figure}[H]
	\centering
	\addOnePanelInFig{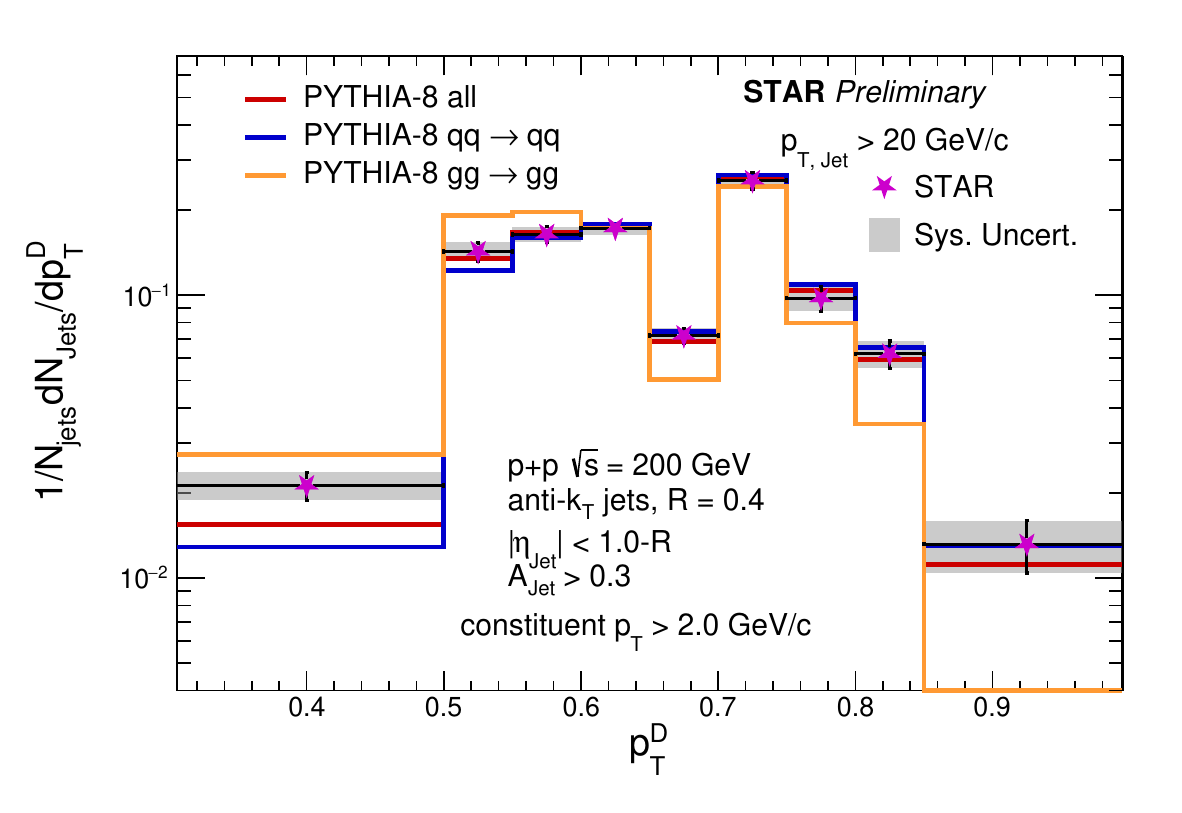}
	\addOnePanelInFig{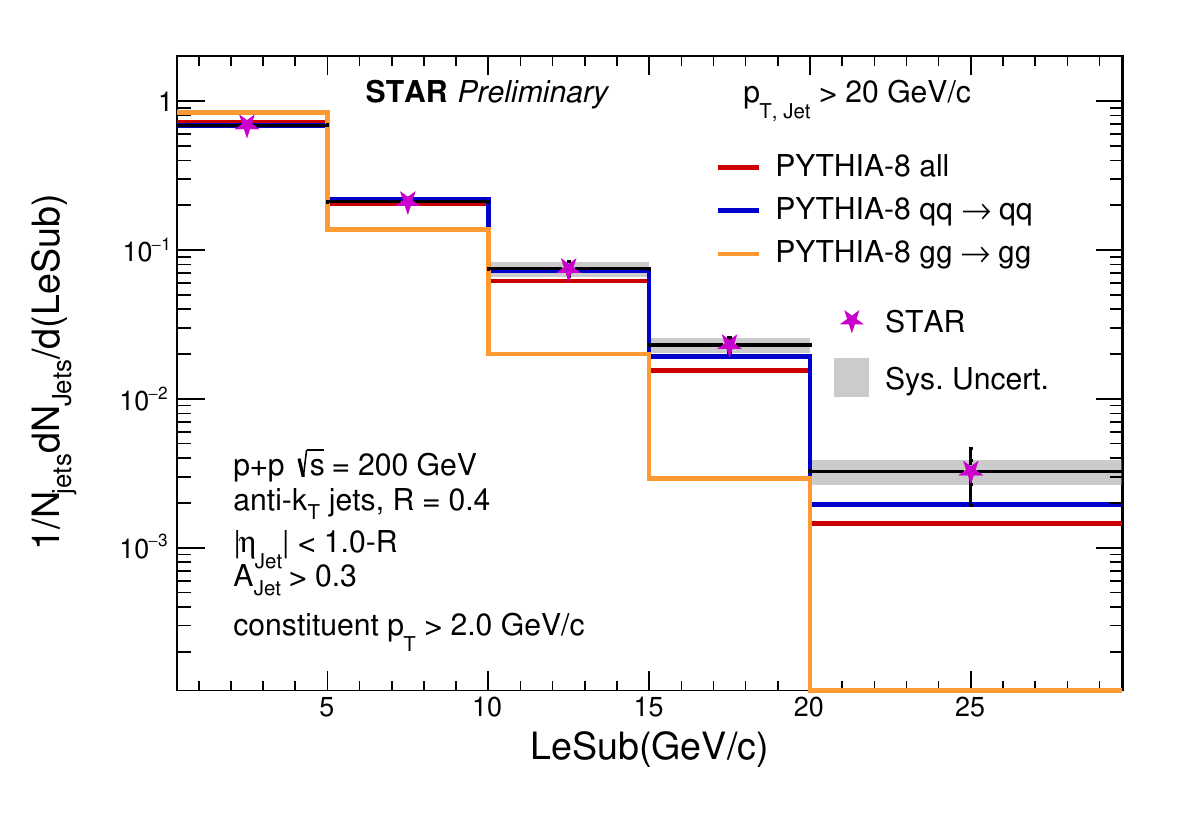}
	\addOnePanelInFig{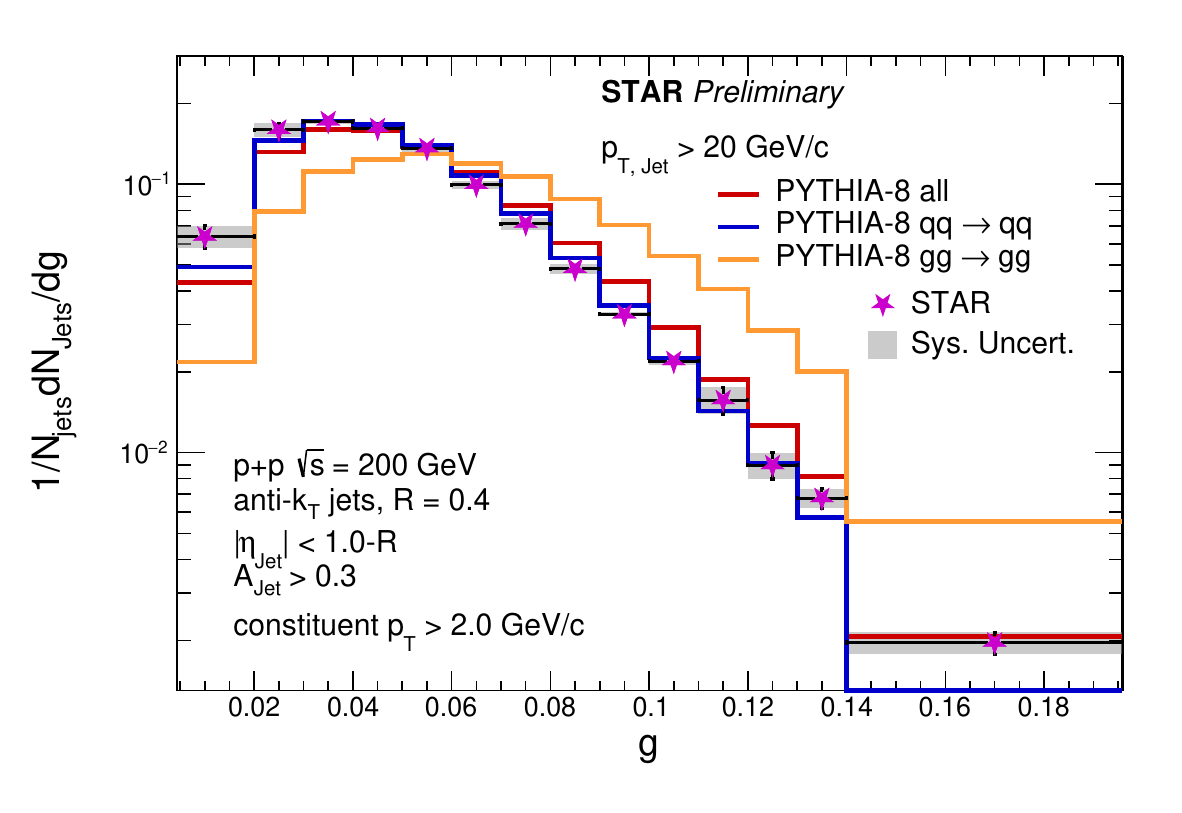}
	\caption{ $\ptd = \sqrt{\lambda_0^2}$ (top-left), \lesub (top-right) and \g~(bottom) distributions for jets with \ptJetGeq{20}. The results are compared to PYTHIA 8 (Detroit) with all hard processes (red), with only \qqToqq processes (blue) and \ggTogg processes (orange).}
	\label{fig:JSOs_QG}
\end{figure}

\begin{figure}[H]
	\centering
	\includegraphics[width = 0.675\linewidth]{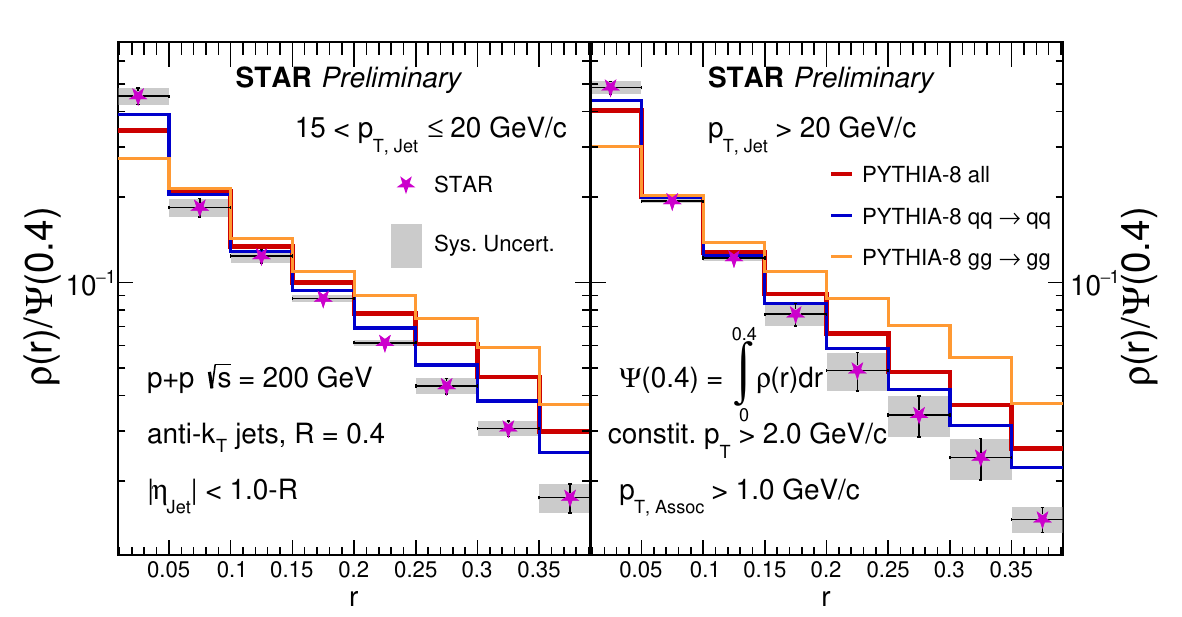}
	\caption{\rhoR~vs $r$ (magenta) for \ptJetRange{15}{20} (left)  and \ptJetGeq{20} (right). The results are compared to PYTHIA 8 (Detroit) with all hard processes (red), with only \qqToqq processes (blue) and \ggTogg processes (orange).}
	\label{fig:RhoR_QG}
\end{figure}

The results are compared to PYTHIA-6 (STAR)~\cite{adkins2019studying}  and PYTHIA-8 Detroit underlying event tune \cite{PhysRevD.105.016011}. All measurements show a good agreement with PYTHIA-6, while PYTHIA-8 is shown to underestimate jets with higher $LeSub$ and lower $g$ values. \rhoR~from PYTHIA-8 underestimates the fraction of jet momentum closer to the jet axis. Figures~\ref{fig:JSOs_QG} and~\ref{fig:RhoR_QG} show STAR data compared to PYTHIA-8 (Detroit) with (a) all hard scatterings, (b) only \qqToqq hard scatterings (quark jets), and (c) only \ggTogg hard scatterings (gluon jets). 

Since gluon jets have softer, more spread-out radiation pattern on average than quark jets \cite{Gallicchio_2011}, they are likely to have lower \ptd, lower \lesub, higher \g~with more momentum (\rhoR) away from the jet-axis. As even quark-jets from PYTHIA-8 (Detroit) show softer fragmentation on average than the STAR data, it is likely that PYTHIA-8 (Detroit) underestimates hard fragmentation of partons.

\section{Conclusions} \label{sect:Conclusions}
First measurements of jet-shape observables \g, \ptd, \lesub~and \rhoR~from STAR using hard-core jets \pp collisions at \sGev~are presented, setting the baseline for heavy-ion collisions to measure the medium-modification at RHIC. With the hard-core jet definition and HT trigger requirement, the sample of jets used here is biased towards hard-fragmented jets. The results show good agreement with PYTHIA-6 (STAR). PYTHIA-8 (Detroit) is shown to underestimate harder-fragmented jets, and needs further tuning of PYTHIA-8's parton shower/hadronization parameters to explain STAR hard-core jets.

\textit{This work is supported by the National Science Foundation under Grant number: 1913624.}

\bibliographystyle{JHEP}	
\bibliography{references}

\end{document}